\documentstyle[twocolumn,aps]{revtex}

\tightenlines
\begin{document}

\newcommand{\beq}{\begin{equation}}
\newcommand{\eeq}{\end{equation}}
\newcommand{\bea}{\begin{eqnarray}}
\newcommand{\eea}{\end{eqnarray}}
\newcommand{\ba}{\begin{array}}
\newcommand{\ea}{\end{array}}
\newcommand{\om}{(\omega )}
\newcommand{\bef}{\begin{figure}}
\newcommand{\eef}{\end{figure}}
\newcommand{\leg}[1]{\caption{\protect\rm{\protect\footnotesize{#1}}}} 

\newcommand{\ew}[1]{\langle{#1}\rangle}
\newcommand{\be}[1]{\mid\!{#1}\!\mid}
\newcommand{\no}{\nonumber}
\newcommand{\etal}{{\em et~al }}
\newcommand{\geff}{g_{\mbox{\it{\scriptsize{eff}}}}} 
\newcommand{\da}[1]{{#1}^\dagger}
\newcommand{\cf}{{\it cf.\/}\ }
\newcommand{\ie}{{\it i.e.\/}\ }
\newcommand{\eg}{{\it e.g.\/}\ }

\title{Evaluating quantum teleportation of coherent states}
\author{Philippe Grangier and Fr\'ed\'eric Grosshans}
\address{Laboratoire Charles Fabry, Institut d'Optique Th\'eorique et Appliqu\'ee, 
F-91403 Orsay, France}

\maketitle

\begin{abstract}

We discuss the criteria for teleporting coherent states
from simple considerations about information exchange
during the teleportation process.




\end{abstract}


\section{Introduction}

Quantum teleportation has emerged in recent years as a major paradigm of 
theoretical \cite{BBC93} and experimental \cite{BPMBBM,FLB98} quantum information. 
The initial approaches using discrete variables \cite{BBC93,BPMBBM} have been extended
to continuous quantum variables \cite{FLB98,BK98,RL98,dd}. Though there is a general agreement
about the main ideas of quantum teleportation, discussions have appeared about
the significance and the evaluation
criteria of real, and thus imperfect, teleportation experiments \cite{BK98,RL98,dd,BK98b,qph}. 

Here our basic requirement for successful teleportation will be that the information 
content of the teleported quantum state is higher than the information 
content of any (classical or quantum) copy of the input state, that may be broadcasted classically,
\ie by using only compatible physical quantities.  
By ``information content" we mean the following : 
the ``verifier" (Victor) is given some classical or quantum information, 
that he tries to compare with the initial  state
that was teleported by Alice, knowing that this state is a pure coherent state $| \alpha \rangle$. 
The quality of this comparison may be characterized by a generalized fidelity,
which is the conditional probability $P(\alpha | I)$ that $| \alpha \rangle$ was sent, given the available 
information $I$. This quantity may eventually be averaged
over the set of initial states $| \alpha \rangle$ \cite{BK98}.
The difference with the ``usual" fidelity is simply that 
$I$ does not need to be a quantum state, it can as well be classical information.

\section{Classical vs quantum fidelity}

Considering first the usual case where the output is a quantum state,
the density matrix of the teleported state
can be expanded on a coherent state basis $| \beta \rangle$, where the probability
to reconstruct the state $| \beta \rangle$ is denoted as $P(\beta)$.
The (standard) fidelity is then simply :
\bea
F_{quant} = \int &d^2 \beta & P(\beta) | \langle \beta | \alpha \rangle |^2 
= \int dx \; dy  P(x,y) \times \no \\ &\exp&(- (x-x_a)^2/4 -  (y-y_a)^2/4)
\eea
where $\alpha = (x_a + i y_a)/2$, $\beta = (x + i y)/2$, and the vacuum noise variance
has been normalized to 1. In a gaussian noise hypothesis, $P(x,y)$
is a normalized gaussian function of $x$ and $y$, centered on the values $x_b$ and $y_b$. 
The $x$ and $y$ variances are the equivalent input noise \cite{PRG94} $N_X^{out}$ and  $N_Y^{out}$ 
in the teleportation process, which are discussed in ref. \cite{qph}. 
One has thus : 
\bea
P(x,y) &=& \frac{1}{2 \pi \sqrt{N_X^{out} N_Y^{out}}} \; \; 
e^{-\frac{(x-x_b)^2}{2 N_X^{out}} - \frac{(y-y_b)^2}{2 N_Y^{out}}}
\eea
We obtain by carrying out the integration : 
\beq
F_{quant}  =
\frac{2}{ \scriptstyle{\sqrt{(2+N_X^{out})(2+N_Y^{out})}}}  \; \; 
e^{ -\frac{(x_a-x_b)^2}{2(2+N_X^{out})} - \frac{(y_a-y_b)^2}{2(2+N_Y^{out})} }
\eeq
The fidelity is thus strongly peaked on the condition $(x_a = x_b, \; y_a = y_b)$, 
which is obtained for unity gain ($g_T=1$)in the teleportation scheme.
Assuming that this condition is satisfied (this is easy to do
in practice \cite{FLB98}), one obtains \cite{FLB98,BK98} :
\beq
F_{quant}^{g_T=1} = \frac{2}{\sqrt{(2+N_X^{out})(2+N_Y^{out})}}
\label{quant}
\eeq
This quantity is clearly relevant for characterizing quantum teleportation,
and can reach the value $F_{quant}^{g_T=1} = 1$ when $N_X^{out} = N_Y^{out} =0$, \ie 
when the teleportation noise is zero. 
For getting a qualitative idea of the classical limit of the teleportation process, 
let us consider the case
where the input beam is split in two equal parts,
and two homodyne measurements shifted by $\pi/2$ are done on each part.
The splitting introduces a vacuum fluctuation mode $v1$.
Then the measured quantities are used to reconstruct the input state,
which introduces another vacuum fluctuations mode $v2$. 
It can be shown (see \eg \cite{qph}) that an optimized measurement procedure will give :
\bea
X_{out} &=& X_{in} + X_{v1} + X_{v2} \nonumber \\
Y_{out} &=& Y_{in} - Y_{v1} + Y_{v2}
\eea
From this equation one gets the classical limits :
\bea
&N_X^{out} = (\Delta X_{v1})^2 + (\Delta X_{v2})^2 = 2& \nonumber \\
&N_Y^{out} = (\Delta Y_{v1})^2 + (\Delta Y_{v2})^2 = 2&  \nonumber \\
&F_{quant}^{g_T=1} = 1/2&
\eea
This corresponds to having twice the shot noise, or using the terminology
of ref. \cite{FLB98}, two ``qduties", one being associated with the 
measurement stage, and the other one with the reconstruction stage.
On the other hand, by using EPR beams \cite{FLB98},
the fluctuations of the two stages are perfectly correlated 
for one quadrature, and anticorrelated for the other one, 
yielding ideally $N_X^{out} = N_Y^{out} =0$.

We consider now the case case where the output is directly obtained from the measurement outcome,
without really reconstructing a quantum state. However, a quite similar reasoning can be applied :
one can guess that the input state is $| \beta \rangle$ with a probability $P_{cl}(\beta)$,
and the conditional probability to obtain the correct answer $| \alpha \rangle$
given the available information $P_{cl}(\beta)$ is :
\bea
F_{class} = \int &d^2 \beta & P_{cl}(\beta) | \langle \beta | \alpha \rangle |^2 
= \int dx \; dy  P_{cl}(x,y) \times \no \\ &\exp&(- (x-x_a)^2/4 -  (y-y_a)^2/4)
\eea
with the same definitions as before. 
The difference with the previous case is that the $x$ and $y$ variances 
are now the noises $N_X^{m}$ and  $N_Y^{m}$ associated
with the measurement, rather than with the full reconstruction procedure. 
By the same calculation as above, one obtains thus  :
\beq
F_{class}^{g_T=1} =  2 \; / \sqrt{(2+N_X^{m})(2+N_Y^{m})}
\eeq
Considering again the case where the input beam is split in two equal parts,
and two homodyne measurements shifted by $\pi/2$ are done on each part,
the classical measurement outcome $X_m$ and $Y_m$ are given by \cite{qph} :
\bea
X_{m} &=& X_{in} + X_{v1}\nonumber \\
Y_{m} &=& Y_{in} - Y_{v1}
\eea
However, contrary to the previous case, the measurement noise must take into account
not only the equivalent input noise
$N_X^{v1} = N_Y^{v1} =1$ which is due to the beamsplitting process, 
but also the noise in the input mode, which is $N_X^{in} = N_Y^{in} =1$
for a coherent state input.
One gets thus again $N_X^{m} = N_Y^{m} = 2$, and therefore  $F_{class}^{g_T=1} =  1/2$.

\section{Discussion}

These two cases correspond to two different views on the teleportation process,
which in some sense are associated to either an ``Heisenberg" (\ie operatorial) or a 
``Schroedinger" (\ie quantum state) picture. 

In the first (Heisenberg-type) view, 
the teleportation process is described by the operatorial equations already given above : 
\bea
X_{out} &=& X_{in} + X_{meas} + X_{rec} \nonumber \\
Y_{out} &=& Y_{in} - Y_{meas} + Y_{rec}
\eea
where $meas$ and $rec$ correspond respectively to the measurement and reconstruction 
procedures. Perfect teleportation is by definition obtained for $X_{out} = X_{in}$, 
$Y_{out} = Y_{in}$. Correspondingly, the noise in the input beam
($X_{in}$, $Y_{in}$) does not contribute to the equivalent input noise
in eq. \ref{quant}. 
The classical limit is thus twice the shot noise (two ``quduties"), and 
perfect teleportation is obtained is when  the measurement and reconstruction 
noises perfectly compensate each other. This point of view is the one
used in ref. \cite{FLB98}, and fits naturally with previous work on 
QND criteria \cite{qph,PRG94}.

In the second (Schroedinger-type) view, the state is first measured. As said above,
the input noise is now relevant, and it is equal to shot-noise for a coherent state.
This input noise plus the beam-splitting noise gives again a classical limit equal
to twice the shot-noise. On the other hand, in this picture 
there is {\it no} extra noise associated to the reconstruction : given a measured
$\beta$, one can exactly reconstruct the coherent state $| \beta \rangle$, 
by using a deterministic translation of the vacuum. 

Some confusions, in particular in the previous version of this note,
may have been due to mixing up these two point of views. 

\section{Conclusion}

Finally, we note that purification procedures \cite{pp1,pp2} or 
recently demonstrated entanglement criteria \cite{PH}
are compatible with the $F=1/2$ limit. However, following
ref. \cite{qph}, we give below  two arguments that question
the meaning of quantum teleportation of coherent states
for small transmission efficiency of the EPR beams. 

First,  the intensity of one EPR 
beam can be measured in order to use that information to reduce the noise
of the second beam  \cite{ens}. 
Then the noise of the corrected beam can be reduced below shot-noise
only when the losses on each beam are less than 50\%. 
This example is closely related to the 
non-separability argument of ref. \cite{qph}, which 
also requires that ``conditional squeezing" can be obtained on one EPR beam,
given a measurement that is done on the other one.
The requirement that the losses on each EPR beam are less than 50\%
is not compatible with the $F=1/2$ boundary, which can tolerate arbitrarily large losses,
but would be a consequence of the requirement $F>2/3$.

Second, a possible use for quantum teleportation is to send a quantum state from Alice to 
Bob for quantum cryptography purposes. In that case, one must worry about the
amount of information which can be eavesdropped during the teleportation process. 
For simplicity, let us consider again a teleportation scheme using EPR beams, 
with a finite degree of squeezing, 
and transmission losses. It is assumed that Eve is able to perfectly eavesdrop the classical channel, 
and that she has full access to the losses along at least one ``transmission arm" of the EPR beam 
(this is  a strong hypothesis, but it is usually done for evaluating the security of standard 
quantum cryptography). 
The simplest solution for Eve is to build her own teleported state, and she will be
successful if this state has an equivalent noise 
smaller than the one achieved by Bob. 
It can be shown simply, and it is physically obvious, that as long as
the EPR channel efficiency $\eta$ is smaller than $1/2$, 
Eve can obtain a teleported copy of the input state which is {\it better}
than the one obtained by Bob. Such low values of $\eta$ can be obtained for values of $F$
larger than $1/2$, but smaller than $2/3$. 

It has been shown in  \cite{pp2} that purifications procedure
can be initiated as soon as $F>1/2$, and may lead to high fidelity values.
However, as long as such procedures are not used,
the above arguments lead to the conclusion that quantum teleportation
with $F<2/3$ may have severe limitations as a quantum communication tool.

{\it Acknowledgements.}
This work was carried out in the framework of the european IST/FET/QIPC project ``QUICOV".
Interesting discussions with R. Jozsa and A. Ekert are acknowledged, as well as
the contribution of I. Cirac for clarifying the previous version of this note.

\end{document}